\documentclass[twocolumn,           % Format : preprint, twocolumn
               showpacs,            % Pacs : showpacs, noshowpacs
               preprintnumbers,     % Preprint: preprintnumbers,
               aps,                 % Society: ...
               prd,          	    % Journal Style : pra, prb, prc, prd, pre,
               letterpaper,             % Size : a4paper, ...
               groupaddress,      % Affiliation (Title) : groupedaddress,
               nofootinbib,         % Footnote: footinbib, nofootinbib
               tightenlines,        % Remove additional spaces in a line
               floats,floatfix      % Floating pictures and tables
               ]{revtex4-1}

\usepackage{graphicx}
\usepackage{latexsym}
\usepackage{amsmath,amssymb}        % amssymb includes amsfonts
\usepackage[draft=false]{hyperref}

\begin{document}

\title{Supersymmetric classical cosmology}

\author{Celia Escamilla-Rivera}
 \email{celia\_escamilla@ehu.es}
 \affiliation{Departamento de F\'isica, Divisi\'on de Ciencias e
   Ingenier\'ias, Campus Le\'on, Universidad de Guanajuato,
   C.P. 37150, Le\'on, Gto., Mexico, \\
   and Fisika Teorikoaren eta Zientziaren Historia Saila, Zientzia eta
   Teknologia Fakultatea, Euskal Herriko Unibertsitatea, 644 Posta
   Kutxatila, 48080, Bilbao, Spain}
 
\author{Octavio Obreg\'on}
 \email{octavio@fisica.ugto.mx}
 \affiliation{Departamento de F\'isica, Divisi\'on de Ciencias e
   Ingenier\'ias, Campus Le\'on, Universidad de Guanajuato,
   C.P. 37150, Le\'on, Gto., Mexico}

\author{L. Arturo Ure\~na-L\'opez}
 \email{lurena@fisica.ugto.mx}
 \affiliation{Departamento de F\'isica, Divisi\'on de Ciencias e
   Ingenier\'ias, Campus Le\'on, Universidad de Guanajuato,
   C.P. 37150, Le\'on, Gto., Mexico}
    
\date{\today}

\begin{abstract}
In this work a supersymmetric cosmological model is analyzed in which
we consider a general superfield action of a homogeneous scalar field
supermultiplet interacting with the scale factor in a supersymmetric
FRW model. There appear fermionic superpartners associated with both
the scale factor and the scalar field, and classical equations of
motion are obtained from the super-Wheeler-DeWitt equation through the
usual WKB method. The resulting supersymmetric Einstein-Klein-Gordon
equations contain extra radiation and stiff matter terms, and we study
their solutions in flat space for different scalar field
potentials. The solutions are compared to the standard case, in
particular those corresponding to the exponential potential, and their
implications for the dynamics of the early Universe are discussed in
turn.
\end{abstract}

         
\maketitle

\section{Introduction}
From the latest observations, we do know that about $95\%$ of matter
in the Universe \cite{Larson:2010gs} is of non-baryonic nature, and
the rest is constituted by radiation, baryons, neutrinos, and all
other particles we understand well in the Standard Model of Particle
Physics. The most successful models until now is the so called Lambda
Cold Dark Matter ($\Lambda$CDM)
model\cite{Olive:2010mh,Bianchi:2010uw,Sapone:2010iz,Linder:2010ks,Colafrancesco:2010zn,Benson:2010de},
which is able to explain and to fit reasonably well all cosmological
observations. 

In the last decades cosmologists have made use of scalar fields in the
description of various aspects of cosmology. They are known in models
of inflation, and more recently in models of dark energy,
see\cite{Copeland:2006wr,Tsujikawa:2010sc} and references therein. But
scalar fields have been considered too for models of dark
matter\cite{Sahni:1999qe,Matos:2000ng}.

This flexibility of scalar field models to describe different
phenomena comes from the properties of the self-interacting scalar
field potential $V(\varphi)$ that is specified for each
model. Currently, there is no underlying principle that uniquely
specifies the potential for the scalar field and many proposals have
been
considered\cite{Alabidi:2010sf,Copeland:2006wr,Liddle:2000cg}. Some
were based in new particle physics and gravitational theories, other
were postulated ad-hoc to obtain the desired evolution. 

On the other hand, the physics required to understand the early
Universe should be necessarily rooted in a theory of quantum
gravity. Futhermore, it would probably be adequate to consider
scenarios where both bosonic and fermionic matter fields would be
present on an equal footing. 

In considering the quantum creation of the Universe we are of course
dealing with the earliest epochs of the Universe's existence, at which
time it is believed that supersymmetry would not yet be broken. The
inclusion of supersymmetry could therefore be vital from the point of
view of physical consistency. 

For these and other physical reasons supersymmetric quantum cosmology
emerged as an active area of research. The first model
proposed\cite{Macias:1987ir} was based on the fact that, shortly after
the invention of supergravity, it was
shown\cite{Teitelboim:1977fs,Tabensky:1977ic} that this theory
provides a natural classical square root equations and their
corresponding Hamiltonians.

A second method later proposed was a superfield formulation, in which
is possible to obtain the corresponding fermionic partners and also
being able to incorporate matter in a simpler
way\cite{Obregon:1996dt,Tkach:1996ph,Obregon:1999wt}. A third method
allows us to define a \textit{square root} of the potential, in the
minisuperspace, of the cosmological model of interest and consequently
operators which square results is the
Hamiltonian\cite{Graham:1991av,Bene:1993ci,Obregon:1993za}.

So, in the same way that we seek a desirable scalar field potential
to explain the evolution (and early times) of the Universe from the
point of view of standard General Relativity, we can reconcile
these requirements along with the ideas of local supersymmetry using
now \textit{superpotentials}. For this purpose we need to model a
supersymmetric quantum cosmological landscape and see what happens now
with the super expansion factor and with the super-scalar fields. It
is then important to find the influence of the "fermionic'' variables
in these superfields and how they would alter the dynamics of usual
cosmological models.

In this work, we consider a Hamiltonian for a homogeneous super scale
factor, which is a supermultiplet (with four components and different
signs) in supergravity $N=2$, and that interacts with a super scalar
field (also a supermultiplet)\cite{Obregon:1999wt}. We shall promote
this Hamiltonian to be an operator, representing the Grassmann
variables by matrices, then by means of the WKB procedure we find two
(independent) classical evolution equations. Those associated with the
scalar field are obtained through Hamilton equations. 

This procedure gives us a modified Einstein-Klein-Gordon (EKG) set of
equations (that we call SUSY-EKG equations) due to the indirect presence
of the "gravitinos'' and the "fermionic'' variables corresponding to
the scalar field, which are inherently contained in each entry of the
supermultiplets. From a phenomenological point of view, the new extra
terms in the model offer different kind of components that behave as
radiation and stiff matter.

The paper is organized as follows. In
Sec.~\ref{sec:susy-ekg-equations}, we outline the procedure that
allows us to define the superfields associated with the expansion
factor and the scalar field so that we can generalize the usual action
of Cosmology. We find the Hamiltonian of the system, which already
contains extra terms depending upon the Grassmanian variables
associated with the scale factor and the scalar field.  

The Grassmanian variables are represented as matrices, and then the
Hamiltonian operator is a matrix itself with four components. We focus
our attention in its two independent components, and apply to them the
usual WKB method to get \emph{classical} equations of motion. Some
solutions are found for the cases of a free scalar field and of a
constant scalar field potential which is negative definite.

Sec.~\ref{sec:susy-cosmology-with} is dedicated to the analysis of the
case in which the scalar field is endowed with an exponential super
potential. We first show that there is an exact scaling solution, in
which all energy terms behave like stiff matter. To have a complete
picture of the solutions, the equations of motion are written as a
dynamical system, and we study its critical points and general
trajectories in the phase space of the resulting variables.

Finally, Sec.~\ref{sec:conclusions} is devoted to conclusions and
comments about the general properties of the SUSY-EKG equations and
their solutions.

%------------------------------------------------------------
\section{The SUSY-EKG Equations for a FRW
  Universe \label{sec:susy-ekg-equations}}
In this section, we describe the main features of SGR cosmology, and
for this we will write the supersymmetric version of the WDW equation
according to the superfield method outlined in the introduction.
 
%------------------------------------------------------------
\subsection{Mathematical background \label{sec:math-backgr-}}
For a homogeneous and isotropic universe, we write the
Friedmann-Robertson-Walker metric as (in units
with $c=1$),
\begin{equation}
  \label{eq:metric}
  ds^{2} = - N(t)dt^{2} + a^{2}(t) \left[ \frac{dr^{2}}{1-kr^{2}} +
    r^{2} d{\Omega}^{2} \right] \, , 
\end{equation} 
where $a(t)$ is the (time-dependent) scale factor, $N(t)$ is the lapse
function, and $k$ is the curvature constant. Then, we can write the
total action representing a (real) scalar field $\phi$ endowed with a
scalar field potential $V(\phi)$, and interacting with the expansion
factor as 
\begin{equation}
  \label{eq:action2}
  S = \frac{6}{8\pi G} \int{ \left( -\frac{a{\dot{a}}^{2}}{2N} +
      \frac{1}{2} kNa \right) dt} + S_{mat}(\Phi) \, . 
\end{equation} 
The equations of motion arising from this action, for $N=1$, are
\begin{subequations}
  \label{eq:motion}
  \begin{eqnarray}
    \dot{H} &=& -\frac{\kappa^2}{6} \dot{\phi}^2 \, , \label{eq:motion1} \\
    \ddot{\phi} &=& -3 H \dot{\phi} - \frac{dV}{d\phi} \,
     , \label{eq:motion2}
  \end{eqnarray}
\end{subequations}
together with the (constraint) Friedmann equation
\begin{equation}
  H^{2} = \frac{\kappa^2}{3} \left( \dot{\phi}^2 + V(\phi) \right) \,
    , \label{eq:motion3}
\end{equation}
where $H \equiv \dot{a}/a$ is the Hubble
parameter, and $\kappa^2 = 8\pi G$. Eqs.~(\ref{eq:motion1}),
(\ref{eq:motion2}), and (\ref{eq:motion3}) are the representative
equations of motion of a FRW universe driven by a scalar field.

%--------------------------------------------------------
We want now to review the procedure that one of us and collaborators
have followed to construct a superfield action for the FRW model
interacting with a (homogeneous) scalar
supermultiplet\cite{Obregon:1999wt}, and from this the
superhamiltonian associated with it.

The most general superfield
action\cite{Obregon:1996dt,Tkach:1996ph,Tkach:1998zw} has the form
\begin{eqnarray}
  \label{eq:superaction}
  S &=& \int 6\left[ - \frac{1}{2\kappa^2}
    \frac{\mathcal{A}}{\mathcal{N}} \mathcal{D}_{\bar{\eta}}
      \mathcal{A} \, \mathcal{D}_\eta \mathcal{A} +
      \frac{\sqrt{k}}{2\kappa^2} \mathcal{A}^2 \right] d\eta \,
    d\bar{\eta} \, dt \nonumber \\
    && + \int \left[ \frac{1}{2} \frac{\mathcal{A}^3}{\mathcal{N}}
      \mathcal{D}_{\bar{\eta}} \Phi \, \mathcal{D}_\eta \Phi - 2
      \mathcal{A}^3 g(\Phi) \right] d\eta \, d\bar{\eta} \, dt \, ,
\end{eqnarray}
where $k=0,1$ denotes flat and closed space, and $\kappa^2 = 8\pi
G_N$, where $G_N$ is Newton's gravitational constant. The units for
the constants and fields in this work are the following:
$[\kappa^2]=\ell^2$, $[\mathcal{N}]= \ell^0$, $[\mathcal{A}]= \ell^1$,
$[\Phi]= \ell^{-1}$, $[g(\Phi)]= \ell^{-3}$, where $\ell$ corresponds
to units of length. Besides, $\mathcal{D}_\eta = \partial_\eta +
i\bar{\eta} \partial_t$ and $\mathcal{D}_{\bar{\eta}} =
- \partial_{\bar{\eta}} - i\eta \partial_t$ are the supercovariant
derivatives of the conformal supersymmetry $N=2$, which has dimension
$[\mathcal{D}_\eta] = [\mathcal{D}_\eta] = \ell^{-1/2}$.

For the one-dimensional gravity superfield
$\mathcal{N}(t,\eta,\bar{\eta})$ ($\mathcal{N} =
\mathcal{N}^\dagger$), we have the following series expansion,
\begin{equation}
  \label{eq:superN}
  \mathcal{N}(t,\eta,\bar{\eta}) = N(t) + i\eta \bar{\psi}^\prime (t)
  + i\bar{\eta} \psi^\prime (t) + \eta \bar{\eta} \mathcal{V}^\prime
  \, ,
\end{equation}
where $N(t)$ is the lapse function, and we have also introduced the
reparametrization $\psi^\prime(t) = N^{1/2}(t) \psi (t)$, and
$\mathcal{V}^\prime (t) = \mathcal{N}(t) \mathcal{V}(t) +
\bar{\psi}(t) \psi (t)$. The Taylor series expansion of the superfield
$\mathcal{A}$ has a similar form,
\begin{equation}
  \label{eq:superA}
  \mathcal{A}(t,\eta,\bar{\eta}) = a(t) + i\eta \bar{\lambda}^\prime
  (t) + i\bar{\eta} \lambda^\prime (t) + \eta \bar{\eta}
  \mathcal{B}^\prime \, ,
\end{equation}
where $a(t)$ is the scale factor, $\lambda^\prime(t) = \kappa
N^{1/2}(t) \lambda (t)$, and $\mathcal{B}^\prime (t) = \kappa N(t)
\mathcal{B}(t) + (1/2) \kappa (\bar{\psi}(t) \lambda(t) - \psi (t)
\bar{\lambda} (t))$. Likewise, the scalar superfield $\Phi(t, \eta,
\bar{\eta})$ may be written as ($\Phi = \Phi^\dagger$),
\begin{equation}
  \label{eq:superPhi}
  \Phi(t,\eta,\bar{\eta}) = \phi(t) + i\eta \bar{\chi}^\prime
  (t) + i\bar{\eta} \chi^\prime (t) + \eta \bar{\eta}
  \mathcal{F}^\prime \, ,
\end{equation}
where $\chi^\prime(t) = N^{1/2}(t) \chi (t)$, and $\mathcal{F}^\prime
(t) = N(t) F(t) + (1/2) (\bar{\psi}(t) \chi(t) - \psi (t) \bar{\chi}
(t))$.

As it was shown in Ref.\cite{Obregon:1996dt}, we now expand the
action~(\ref{eq:superaction}) in terms of the superfield
components~(\ref{eq:superN}), (\ref{eq:superA}),
and~(\ref{eq:superPhi}), and integrate over the Grassmann complex
coordinates $\eta$ and $\bar{\eta}$. If we redefine
\begin{equation}
  \lambda (t) \to \frac{1}{3} a^{-1/2}(t) \lambda (t) \, , \quad \chi
  (t) \to a^{-3/2}(t) \chi (t) \, ,
\end{equation}
it is possible to find the Lagrangian, and from it the
superHamiltonian can be constructed, namely,
\begin{eqnarray}
  \mathcal{H} &=& -\frac{\kappa^{2}}{12} a^{1/2} \Pi_{a} a^{1/2} \Pi_{a} -
  \frac{3ka}{\kappa^{2}} - \frac{1}{6} \frac{\sqrt{k}}{a} \left[
    \bar{\lambda},\lambda \right] + \frac{\Pi^{2}_{\varphi}}{2a^{3}}
  \nonumber \\ 
  && - \frac{i\kappa}{4a^{3}} \Pi_{\varphi} \left( \left[
      \bar{\lambda},\chi \right] + \left[ \lambda,\bar{\chi} \right]
  \right) - \frac{\kappa^{2}}{16a^{3}} \left[ \bar{\lambda},\lambda
  \right] \left[ \bar{\chi},\chi \right] \nonumber \\ 
  && + \frac{3\sqrt{k}}{4a} \left[ \bar{\chi},\chi \right] +
  \frac{\kappa^{2}}{2}g \left( \varphi \right) \left[
    \bar{\lambda},\lambda \right] + 6\sqrt{k}g \left( \varphi
  \right) a^{2} \nonumber \\ 
  && + a^{3} V \left( \varphi \right) + \frac{3}{4} \kappa^{2}g
  \left( \varphi \right) \left[ \bar{\chi},\chi \right] +
  \frac{\partial^{2}g \left( \varphi \right)}{\partial \varphi^{2}}
  \left[ \bar{\chi},\chi \right] \nonumber \\ 
  && + \frac{\kappa}{2} \frac{\partial g \left( \varphi \right)}{\partial
    \varphi} \left( \left[ \bar{\lambda},\chi \right] - \left[
      \lambda,\bar{\chi} \right] \right) \, . \label{hamiltonian susy}
\end{eqnarray}
where the scalar field potential reads
\begin{equation}
\label{potential}
V ( \varphi ) = 2 \left( \frac{\partial g(\varphi)}{\partial \varphi}
\right)^{2} - 3\kappa^{2}g^{2}(\varphi) \, .
\end{equation}

Notice that, in general, the scalar potential~(\ref{potential}) is not
positive semi-definite. The relevant term in Eq.~(\ref{potential}) is
$g(\varphi)$, which is related to the superpotential and whose form
shall be chosen appropriately for the cosmological model under study.

In the quantum (canonical) formalism the Grassmanian variables
$\lambda$, $\bar{\lambda}$, $\chi$, and $\bar{\chi}$, --- by the
anticommutators as 
\begin{equation}
  \label{clifford}
  \left\{ \lambda,\bar{\lambda} \right\} = -\frac{3}{2} \, , \quad
  \left\{ \chi,\bar{\chi} \right\} = 1 \, ,
\end{equation}
and they can be considered as generators of the Clifford algebra, as
well as the commutators
\begin{equation}
  \label{eq:commutators}
  \left[ a,\Pi_a \right] = -i \, , \quad \left[ \phi,\Pi_\phi \right]
  = -i \, .
\end{equation}

We can choose a matrix representation for the "fermionic'' operators
$\lambda$, $\bar{\lambda}$, $\chi$, and $\bar{\chi}$, in the form of a
tensorial products of $2\times 2$ matrices,
\begin{subequations}
\label{eq:operators-matrix}
  \begin{eqnarray}
    \lambda = \sqrt{\frac{3}{2}} \sigma_{-} \otimes 1 \, , &\quad&
    \bar{\lambda} = - \sqrt{\frac{3}{2}} \sigma_{+} \otimes 1 \, ,  \\
    \chi = \sigma_{3} \otimes \sigma_{-} \, , &\quad& \bar{\chi} =
    \sigma_{3} \otimes \sigma_{+} \, , 
  \end{eqnarray}   
\end{subequations}
where $\sigma_{\pm} = (\sigma_{1} \pm i\sigma_{2})/2$, $\sigma_{1}$,
$\sigma_{2}$, and $\sigma_{3}$ are Pauli matrices. 

%---------------------------------------------------
\subsection{The classical landscape \label{sec:classical-landscape-}}
As we have already mentioned in the introduction, our objective is now
to construct the classical equations that corresponds to the
Hamiltonian~(\ref{hamiltonian susy}).

First, we promote it to an operator $\hat{\mathcal{H}}$ by realizing
the "fermionic" variables as the matrices~(\ref{eq:operators-matrix})
and, as usual, $\Pi_a = i\partial_a$ and $\Pi_\phi =
i \partial_\phi$. By these means, we will get a quantum Hamiltonian
operator that should fulfill $\hat{\mathcal{H}} |\Psi \rangle =
0$. Because the matrices in~(\ref{eq:operators-matrix}) are $4 \times
4$, the wave function $\Psi$ will have four components.  

It can be shown\cite{Obregon:1999wt} that the components $\Psi_1$ and
$\Psi_4$ satisfy independent equations, whereas $\Psi_2$ and $\Psi_3$
appear coupled in the other two differential equations. In this work
we focus our attention in the former case, and apply separately to
$\Psi_1$ and $\Psi_4$ the WKB method, so that
\begin{equation}
  \label{eq:WBKphi}
  \Psi = e^{\left( S_{a} + S_{\varphi} \right)} \, .
\end{equation}

With this we shall find the classical equations of motion associated
to the components $\Psi_1$ and $\Psi_4$, which also correspond to the
classical Hamiltonian; from this, classical equations can be obtained
for the scalar field $\varphi$. Thus, the classical equations of
motion are
\begin{subequations}
\label{eq:SUSY-EKGmov}
\begin{eqnarray}
  \ddot{\varphi} &=& - 3 H \dot{\varphi} - \frac{\partial V}{\partial
    \varphi} \pm \frac{3}{2} \frac{\kappa^2}{a^{3}} \frac{\partial
    g(\varphi)}{\partial \varphi} - 6 \frac{\sqrt{k}}{a}
  \frac{\partial g(\varphi)}{\partial \varphi} \nonumber \\
  && \mp \frac{1}{a^{3}} \frac{\partial^{3} g(\varphi)}{\partial
    \varphi^{3}} \, ,  \label{eq:SUSY-EKGmov1} \\
  H^2 &=& \frac{\kappa^2 \dot\varphi^{2}}{6} + \frac{\kappa^2}{3}
  V(\varphi) - \frac{k}{a^{2}} \pm \frac{\kappa^2 \sqrt{k}}{3 a^{4}} +
  \frac{\kappa^4}{32 a^{6}} \nonumber \\
  && \mp \frac{\kappa^4}{2 a^{3}} g(\varphi) + 2 \frac{\kappa^2
    \sqrt{k}}{a} g(\varphi) \pm \frac{\kappa^2}{3 a^{3}}
  \frac{\partial^{2} g(\varphi)}{{\partial \varphi}^{2}} \,
  . \label{eq:SUSY-EKGmov2}
\end{eqnarray}
\end{subequations}
The upper (lower) sign in Eqs.~(\ref{eq:SUSY-EKGmov}) corresponds to
the (quantum) equation for $\Psi_1$ ($\Psi_4$).

We have now SUSY-EKG classical equations that, due to the presence of
variables $\lambda$ and $\bar{\lambda}$ associated with the
"gravitino" and the "fermionic" variables $\chi$ and $\bar{\chi}$
associated with the scalar field, considerably differ from the
standard EKG equations~(\ref{eq:motion}) of General Relativity.

There are extra terms, due to supersymmetry, behaving like
radiation ($a^{-4}$) and stiff matter ($a^{-6}$), which should be
dominant at very early times, whereas other terms show a combination
of scale factor powers mediated by the presence of the superpotential
$g(\phi)$ and its derivatives.

\subsection{Simple classical examples \label{sec:simple-class-exampl}}
\label{sec:simple-classical}
As a first instance of a solution, we will consider the case
$g(\varphi)=0$, which also corresponds to a null potential,
$V(\varphi) = 0$. We also set the curvature term $k=0$. The solutions
of Eqs.~(\ref{eq:SUSY-EKGmov}) are
\begin{eqnarray}
  \label{eq:free}
  \dot{\varphi}(t) &=& \dot{\varphi}_{0} (a_0/a)^3 \, , \\
  a^3(t) &=& a^3_0 + 3 \left( \frac{\kappa^2 \dot{\varphi}^2_0
      a^6_0}{6} + \frac{\kappa^4}{32} \right)^{1/2} (t-t_0) \, ,
\end{eqnarray}
where $t_0$, $\dot{\varphi}_0$, and $a_0$ are integration
constants. The whole solution corresponds to stiff matter, and is
practically the same as in the standard case because the $g$-terms
dissappear from everywhere.

Another more interesting case is that with a constant superpotential,
$g(\varphi) = g_{0}$, that corresponds to a constant and negative
definite scalar field potential, $V = -3\kappa^2 g_{0}^2$. The scalar
field potential is then an effective cosmological constant which is
negative definite.

As in the previous case of the free scalar field, the case is
simplified because the derivatives of the superpotential disappear
from the SUSY-KG equation~(\ref{eq:SUSY-EKGmov1}), though not from the
SUSY-Friedmann equation~(\ref{eq:SUSY-EKGmov2}). The cosmological
solutions can be expressed as
\begin{subequations}
  \begin{eqnarray}
    \dot{\varphi}(t) &=& \dot{\varphi}_{0} (a_0/a)^3 \,
    , \label{SUSY-EKG solution1} \\
    a^3(t) &=& \frac{a_0}{4g_0} \left[ \left(
        \sqrt{\frac{8{\dot{\varphi}^2_0 a^6_0}^{2}}{3\kappa^{2}} +
          \frac{9}{8}} \right) \sin \left[3{\kappa}^{2} {g}_{0}
        (t-t_0) \right] \mp 1 \right] \, , \label{SUSY-EKG solution2}   
  \end{eqnarray}
\end{subequations}
where again $\dot{\varphi}_0$ and $a_0$ are integration constants. 

Because the scale factor is a positive quantity, the only
acceptable solution is when the amplitude of the sinus function is
less or equal to one. It is then clear that the scale factor has a
periodic solution in which $a_0$ is the amplitude at maximum
expansion. In other words, Eq.~(\ref{SUSY-EKG solution2}) represents
an oscillatory Universe. 

%----------------------------------------------------
\section{SUSY Cosmology with an exponential
  superpotential \label{sec:susy-cosmology-with}}
The possible cosmological roles of exponential potentials in scalar
field models have been  thoroughly investigated in the specialized literature
\cite{Wetterich:1987fm,Ratra:1987rm,Chimento:1995da,Ferreira:1997au,Ferreira:1997hj,Copeland:1997et,Barreiro:1999zs,PhysRevD.62.023517,Liddle:1998jc,Billyard:1998hv,vandenHoogen:1999qq,Neupane:2003cs,UrenaLopez:2005zd},
see also\cite{Liddle:2000cg,Copeland:2006wr,Tsujikawa:2010sc}, almost
always as a means for driving a period of cosmological inflation, but
also as possible candidates for dark matter and dark energy.

Scalar field cosmologies with an exponential potentials are, as compared
to others, mathematically simple, and their solutions have many
interesting features. For the purposes of this work, we only mention
the possibility of having inflationary solutions and the appearance of
the so-called scaling solutions, which are nicely illustrated in, for
instance, the dynamical system study presented in\cite{Liddle:1998jc}.

The inflationary solution for exponential potential is the simple
power law inflation\cite{Liddle:2000cg}, which however never ends and
needs modifications to provide a graceful exit towards a Hot Big Bang
model. On the other hand, the scaling solution arises whenever the
scalar field is accompanied by another matter fields, so that both
fields evolve with a fixed ratio of their energy densities, see for
instance\cite{Chimento:1995da,Ferreira:1997au,Copeland:1997et,Barreiro:1999zs,Liddle:1998jc,PhysRevD.62.023517,UrenaLopez:2005zd}.

In this section we explore in detail the type of solutions permitted
by our (classical) SUSY cosmological model when the scalar field is
endowed with an exponential potential. Our main interest will be to
find inflationary and scaling solutions. Even though we are not
considering extra matter fields apart from the scalar field, the new
terms in Eqs.~(\ref{eq:SUSY-EKGmov}) will play the role of companion
fields which should impose a non-trivial behaviour upon the field
$\varphi$.

Let us consider the following superpotential and potential,
respectively,
\begin{subequations}
  \label{eq:gSUSY}
  \begin{eqnarray}
    g(\varphi) &=& g_{0} \, e^{-\lambda \kappa \varphi/2} \,
    , \label{eq:gSUSY1} \\
    V(\varphi) &=& V_0 e^{-\lambda \kappa \varphi} \, , \quad V_0
    \equiv \frac{\kappa^2 g^2_0}{2} \left( \lambda^{2} - 6
    \right) \label{eq:gSUSY2} \, ,   
  \end{eqnarray}
\end{subequations}
where the potential parameters were chosen to ease their comparison
with the standard case; notice that in order to avoid a negative
definite potential we should impose the condition $\lambda >
\sqrt{6}$. The equations of motion~(\ref{eq:SUSY-EKGmov}) with an
exponential superpotential explicitly read
\begin{subequations}
\label{eq:exponential}
  \begin{eqnarray}
    \ddot{\varphi} &=& - 3\frac{\dot{a}}{a} \dot{\varphi} + \lambda
    \kappa V \pm (\lambda^2 - 6) \frac{\lambda \kappa^3 g}{8a^{3}} \,
    , \label{eq:exponentiala} \\
    H^2 &=& \frac{\kappa^2}{6} \dot{\varphi}^2 + \frac{\kappa^2}{3} V
    + \frac{\kappa^4}{32 a^{6}} \pm (\lambda^2 - 6) \frac{\kappa^4
      g}{12a^3} \, . \label{eq:exponentialb}
  \end{eqnarray} 
\end{subequations}

%-----------------------------------------------------
\subsection{Exact SUSY scaling
  solution \label{sec:exact-susy-scaling}}
We present here a first (exact) solution of
Eqs.~(\ref{eq:exponential}) that we shall call \emph{scaling
  solution}, because of its resemblance with the scaling behavior
exponential potentials show in standard
cosmology\cite{Copeland:1997et,Billyard:1998hv,vandenHoogen:1999qq}.

It can be noticed that there is a stiff matter term in
Eq.~(\ref{eq:exponentialb}), and that the superpotential $g$ appears
accompanied by factor $a^{-3}$. Thus, one can foresee that there must
be a stiff matter solution of the equations of motion, so that $a \sim
t^{1/3}$, as long as $g \sim a^{-3}$ and $V \sim a^{-6}$. It can be
shown, just by direct substitution in Eqs.~(\ref{eq:exponential}),
that the exact scaling solution is
\begin{subequations}
  \begin{eqnarray}
    a(t) &=& a_0 (t/t_0)^{1/3} \, , \label{solution scaling1} \\
    \kappa \varphi(t) &=& \frac{2}{\lambda} \ln (t/t_0) \,
    , \label{solution scaling2} \\
    g(t) &=& g_0 (t_0/t) \, , \quad g_0 = \mp a^{-3}_0 \, ,
  \end{eqnarray}
\end{subequations}
where $a_0$ is an appropriate constant to accomplish
Eqs.~(\ref{eq:exponential})\cite{EscamillaRivera:2010zz}.

The scaling solution corresponds to stiff fluid matter, as revealed
by the power law behaviour of the scale factor in Eq.~(\ref{solution
  scaling1}); this is probably not surprising, because we have already
noticed the presence of a stiff-term in the SUSY Friedmann
equation~(\ref{eq:SUSY-EKGmov2}). This solution is not
inflationary, but its existence indicates its possible importance in
the early dynamics of the models considered here.

%-------------------------------------------------
\subsection{Dynamical System Structure \label{sec:dynam-syst-struct}}
Our next step is to study the evolution of our SUSY classical model in
which the scalar field $\varphi$ is endowed with an exponential
potential. As in the standard case, it is possible to perform a
dynamical study of the cosmological model so that its physically
relevant solutions are easily unveiled. 

In order to construct a dynamical system for our cosmological model,
we follow Ref.\cite{Copeland:1997et}, see
also\cite{PhysRevD.62.023517,Coley:1999uh,vandenHoogen:1999qq,UrenaLopez:2005zd}. One first step is to
introduce a set of conveniently chosen variables which may allow the
rewriting of the evolution equations as an autonomous phase system
subject to a constraint arising from the Friedmann equation. 

We choose the following variables,
\begin{equation}
  \label{eq:variables}
  x \equiv \frac{\kappa \dot{\varphi}}{\sqrt{6}H} \, , \quad y \equiv
  \frac{\kappa \sqrt{V}}{\sqrt{3} H} \, , \quad z \equiv
  \frac{\kappa^2}{\sqrt{32} a^{3} H} \, ,
\end{equation}
which render the Friedmann equation as
\begin{equation}
  \label{Fconstriction}
  F(x,y,z) := x^2 + y^2 + z^2 \pm 2\sqrt{(\lambda^2 -6)/3} \, yz = 1
  \, .
\end{equation}
The constraint equation~(\ref{Fconstriction}) follows from
Eq.~(\ref{eq:exponentialb}), and we see that variable $z$ plays the
role of an extra fluid term which, contrary to the standard case, see
Ref.\cite{Copeland:1997et}, is not trivially coupled to the scalar
field variables.

We shall restrict ourselves to the part of the phase space that makes
physical sense, and this is the range $0 \leq |x|,y,z < \infty$, which
is the part that corresponds to expanding universes only. Combining
expressions~(\ref{eq:exponential}) and~(\ref{eq:variables}), the
equations of motion read
\begin{subequations}
\label{evolution eqs exp1}
\begin{eqnarray}
  x^\prime &=& -3x - \frac{\dot{H}}{H^{2}} x + \sqrt{\frac{3}{2}}
  \lambda y^2 \pm \frac{\lambda \sqrt{\lambda^2 - 6}}{\sqrt{2}} yz \, ,
  \\
  y^\prime &=& \sqrt{\frac{3}{2}} \lambda xy - \frac{\dot{H}}{H^{2}} y \, ,
  \\
  z^\prime &=& -3z - \frac{\dot{H}}{H^{2}} z \, ,
\end{eqnarray}
\end{subequations}
where
\begin{eqnarray}
  \frac{\dot{H}}{H^{2}} &=& -3x^2 - 3z^2 \mp \sqrt{3(\lambda^2 -6)} \,
  y z \, .
\end{eqnarray}
Here primes denote derivative with respect to the logarithm of
the scale factor, $N=\ln (a)$. The evolution of phase space variables
$x$, $y$, and $z$ takes place only on the constraint surface
described by Eq.~(\ref{Fconstriction}).

Notice that there is a symmetry in Eqs.~(\ref{evolution eqs exp1})
with respect to the double sign $(\pm)$ and variable $y$. The case
with sign $(-)$ can be obtained from the case with $(+)$ if we change
$y \to -y$; and vice versa. As we shall see below, we will only study
the case with the lower signs since only for them is that we can
obtain positive results for $y$.

\begin{figure*}[htp]
\includegraphics[width=7.7cm]{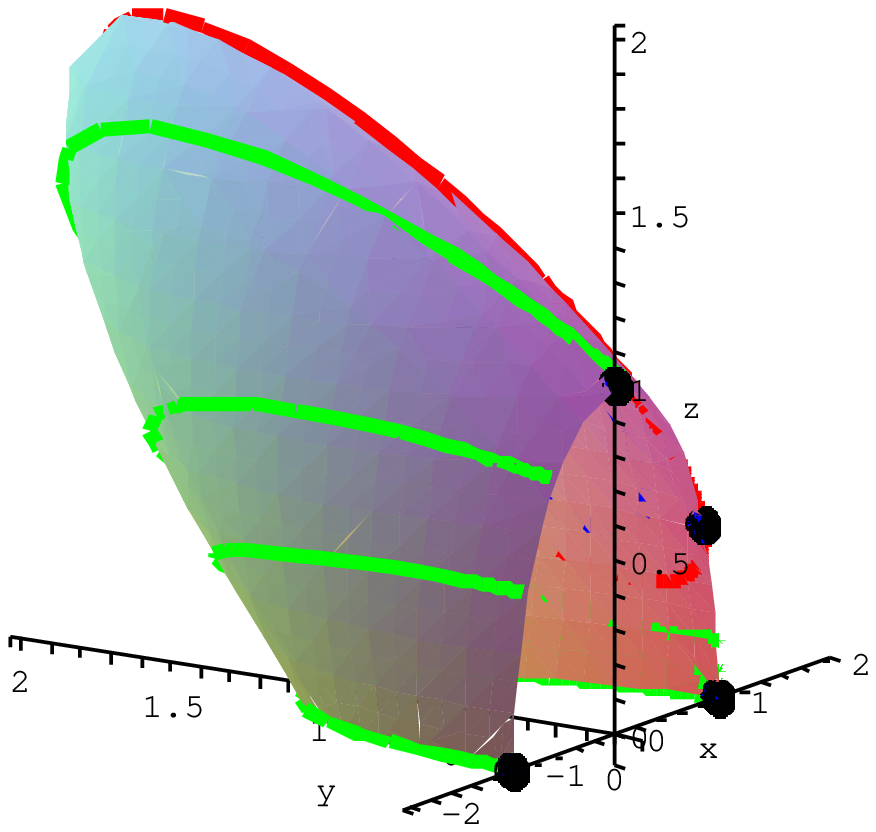}
\includegraphics[width=7.7cm]{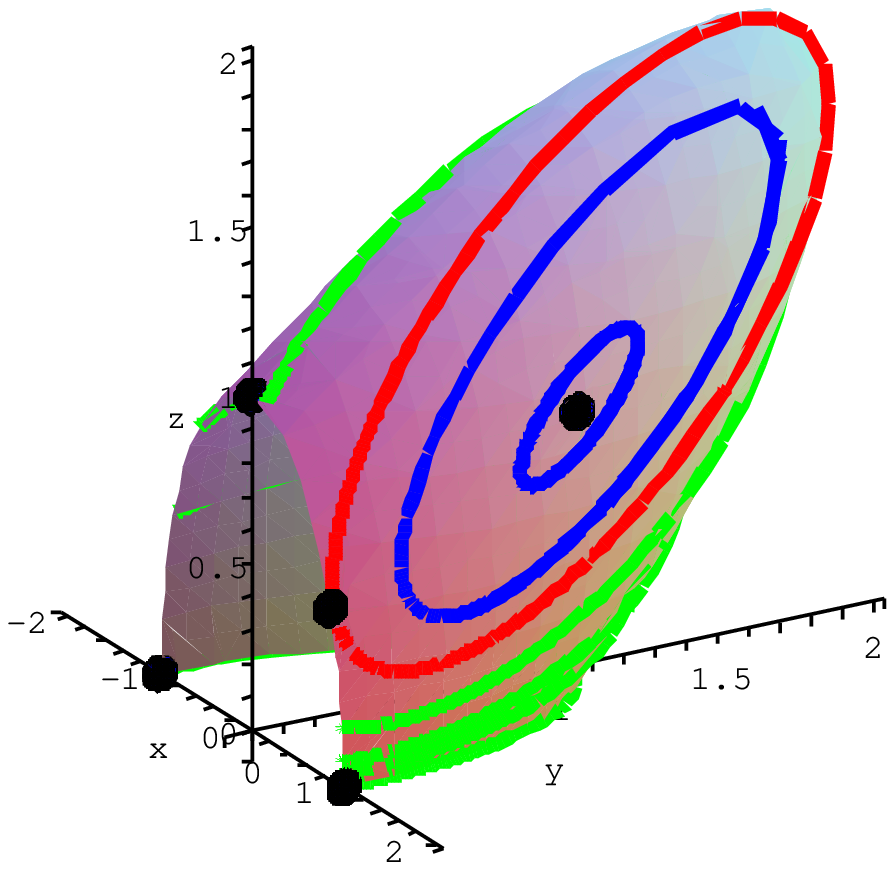}
\caption{\label{fig:fig1} Different views of the phase space of the
  dynamical system~(\ref{evolution eqs exp1}) for the particular value
  $\lambda = 2.85$, so that $6 < \lambda^2 < 9$; the 3-d surface
  represents the constraint $F(x,y,z)=1$, see
  Eq.~(\ref{Fconstriction}), whereas the curves are solutions of the
  dynamical system for diverse initial conditions. The dots denote the
  critical points A, B, C and E shown in Table~\ref{tab:critical}, and
  the trajectories reveal their stability properties as described in
  the text. In particular, notice the largest (red) loop that encircles
  point E (see figure on the right): it is a homoclinic trajectory
  that departs from and arrives to the same critical point.}
\end{figure*}

\begin{figure*}[htp]
\includegraphics[width=7.7cm]{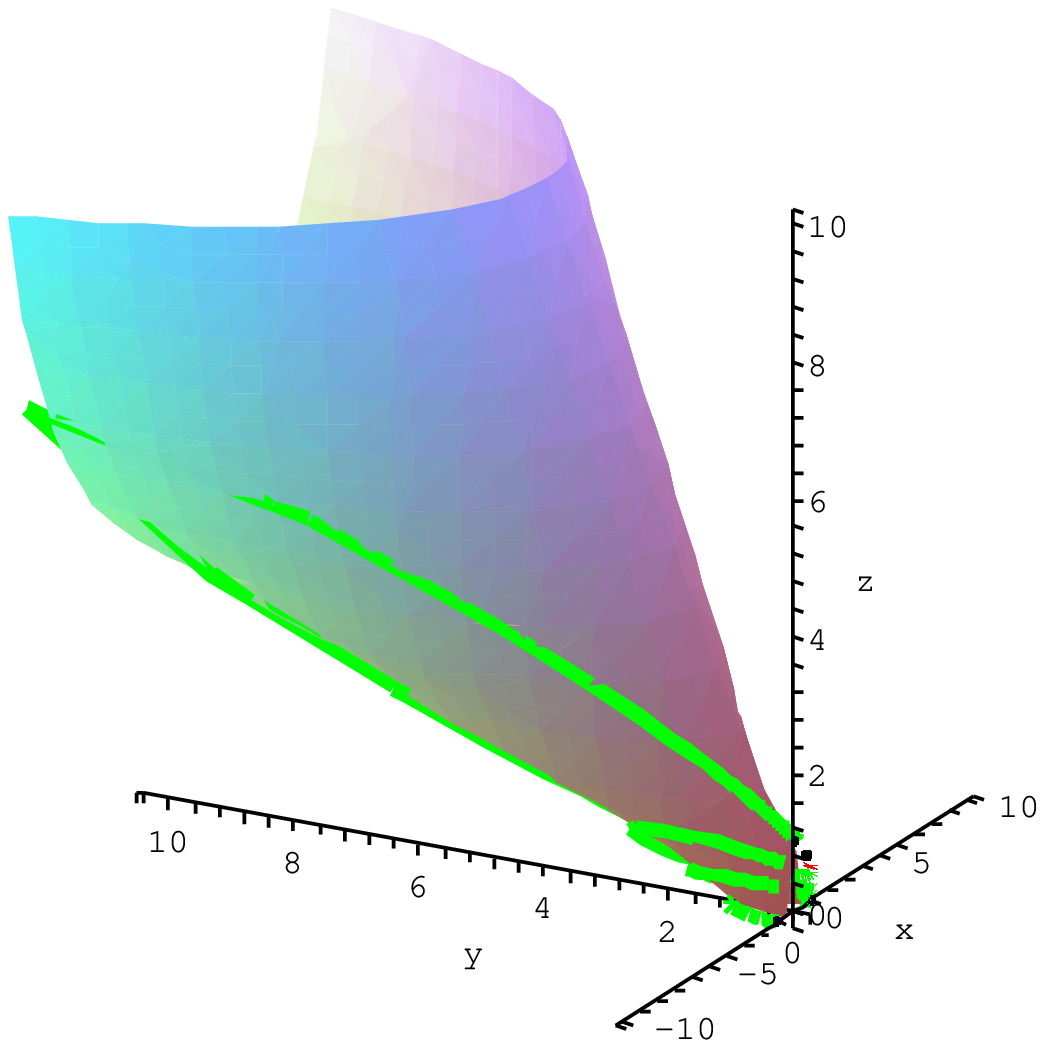}
\includegraphics[width=7.7cm]{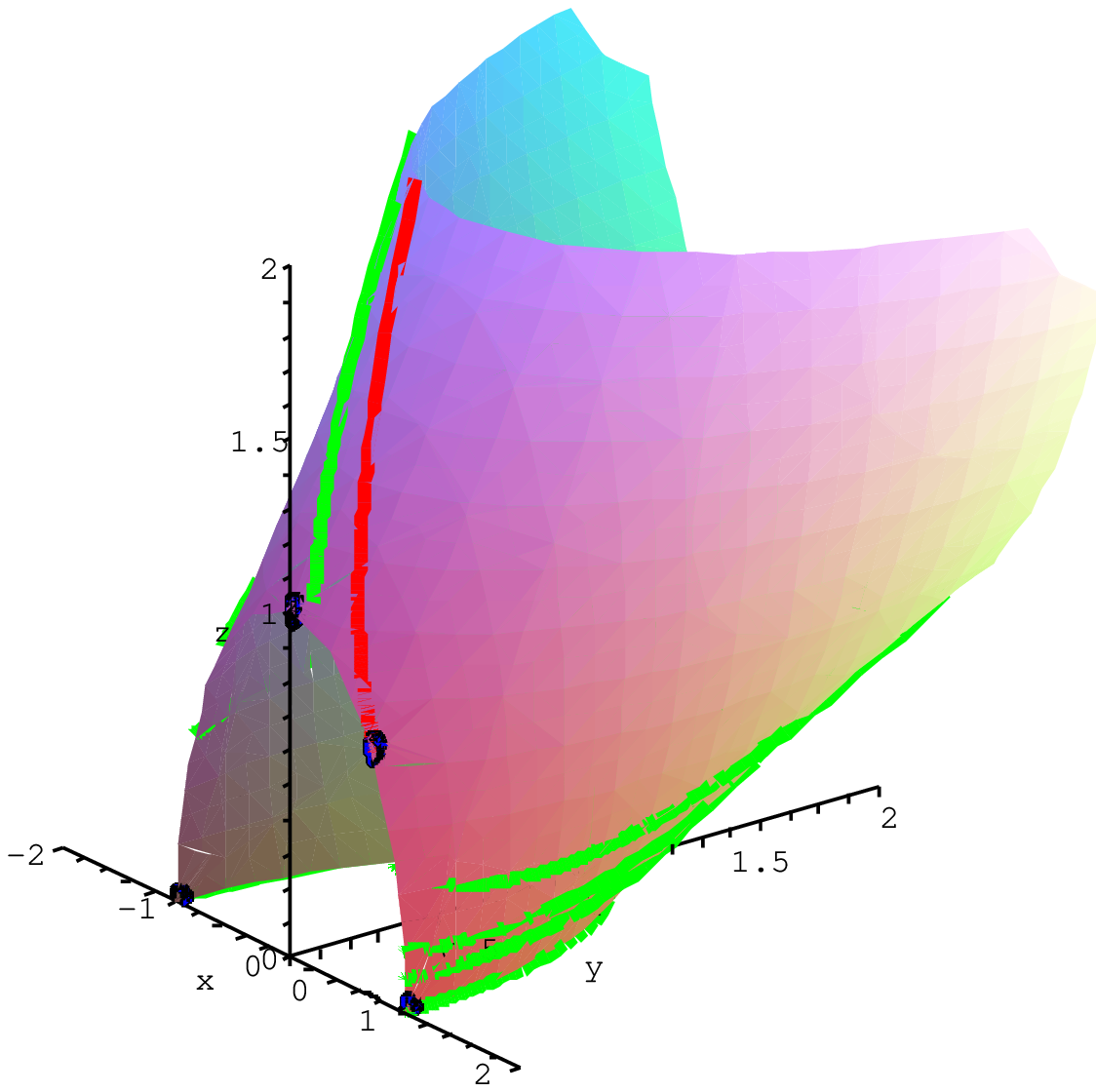}
\caption{\label{fig:fig2} Different views of the phase space of the
  dynamical system~(\ref{evolution eqs exp1}) for the particular value
  $\lambda = 3.5$, so that now $9 < \lambda^2$; the 3-d surface
  represents the constraint $F(x,y,z)=1$, see
  Eq.~(\ref{Fconstriction}), whereas the curves are solutions of the
  dynamical system for diverse initial conditions. The dots denote the
  critical points A, B, and C shown in Table~\ref{tab:critical}, and
  the trajectories reveal their stability properties as described in
  the text. Point E does not exist in this case. As in
  Fig.~\ref{fig:fig1}, there also exists the
  (red) homoclinic trajectory that departs from and arrives to the same
  critical point located at
  $(\sqrt{6}/\lambda,0,\sqrt{1-6/\lambda^2})$. However, we were not
  able to show the complete closed trajectory due to numerical
  limitations.}
\end{figure*}

We show in Table~\ref{tab:critical} the critical points of the
dynamical system~(\ref{evolution eqs exp1}) and their (linear)
stability properties.
\begin{table*}[htp]
\caption{\label{tab:critical} Critical points $(x_\ast,y_\ast,z_\ast)$
  of the dynamical system~(\ref{evolution eqs exp1}) that represents
  the classical supercosmology of a scalar field endowed with an
  exponential superpotential. See also Fig.~\ref{fig:fig1} for a
  graphical representation of the phase space.}
\begin{ruledtabular}
\begin{tabular}{c c c c c c}
Label & $x_\ast$  & $y_\ast$  & $z_\ast$  & Existence  & Stability \\ 
\hline
A     & $0$  & $0$  & $1$  & $\forall \lambda$  & Unstable \\ 

B     & $-1$   & $0$  &$0$  & $\forall \lambda$  & Unstable \\ 
      & $1$   & $0$  &$0$  & $\forall \lambda$  & Unstable for
$\lambda^2 < 6$ \\ 
      & & & & & Saddle for $\lambda^2 > 6$ \\
C     & $-\sqrt{1-z^2_\ast}$  & $0$ & $z_\ast$  &  $\forall \lambda
\, , \; z_\ast < 1$ & Unstable \\ 
     & $\sqrt{1-z^2_\ast}$  & $0$ & $z_\ast$  &  $\forall \lambda
\, , \; z_\ast < 1$ & Unstable for $\sqrt{1 - 6/\lambda^2} < z < 1$ \\ 
      & & & & & Saddle for $0 < z < \sqrt{1 - 6/\lambda^2}$ \\
D     & $\lambda/\sqrt{6}$  & $\sqrt{1-\lambda^2/6}$  & $0$ &
$\lambda^2 < 6$   & Stable node \\ 
%& & & & & Saddle for $\lambda^2 < 6$ \\
E     & $\sqrt{6}/\lambda$ & $(\lambda^2 -6)/[\lambda
\sqrt{9 -\lambda^2}]$ & $\sqrt{3(\lambda^2 -6)}/[\lambda \sqrt{9
  -\lambda^2}]$ & $6 < \lambda^2 < 9$ & Stable centre \\
%    &       &         &         &           & Stable spiral for 
\end{tabular}
\end{ruledtabular}
\end{table*}

There are five critical points, in close similarity to the standard
case, whose main features are described next.
\begin{itemize}
\item Stiff matter domination. The potential variable is null, $y=0$,
  and then the dynamical system is equivalent to the standard case of
  stiff fluid matter ($a^{-6}$) plus a free scalar field ($\dot{\phi}
  \sim a^{-3}$), so that $x^2 + z^2 =1$. Particular (well known) cases
  are:
  \begin{itemize}
  \item Point A, stiff fluid domination. The scalar field variables $x$
    and~$y$ are both null, and then this point represents the
    complete domination of the stiff matter term, $z=1$. This solution
    exists also in the standard cosmological case.
    
  \item Points B, kinetic domination. They represent the domination of
    the scalar field's kinetic energy, $x = \pm 1$, and $y=0=z$. This
    solution exists also in the standard cosmological case.

  \item Points C, joint kinetic and stiff domination. This is a
    (restricted) scaling solution of stiff nature which is satisfied
    by all points inside a unitary circle on the plane $y=0$. This set
    of solutions exists also in the standard cosmological case, but it
    was missed in the analysis of Ref.\cite{Copeland:1997et}: it is
    the line segment $x = [-1,1]$ at $y=0$. Notice that points A (B)
    can be seen as extreme C-points as $z \to 1$ ($z \to 0$).
  \end{itemize}

\item Point D, scalar field domination. It is the coexistence
  of the (scalar) kinetic and potential energies, $x^2 + y^2 = 1$, and
  then the point is located in the unitary circumpherence on the
  plane $z=0$. Notice, however, that the existence of this point
  requires $\lambda^2 < 6$, which is in contradiction with our earlier
  assumption that $\lambda^2 > 6$, see the scalar field potential
  defined through Eq.~(\ref{eq:gSUSY2}).

\item Point E, scaling solution. This point corresponds to the scaling
  solution in Sec.~\ref{sec:exact-susy-scaling}, and represents the
  coexistence of all energy terms in the equations of motion. It
  should be noticed that, contrary to the present work, in the
  standard cosmological case the scaling solution in the presence of
  stiff fluid matter necessarily requires $y=0$, see Table~I in
  Ref.\cite{Copeland:1997et}.
\end{itemize}

The stability of the points is investigated through linear
perturbations around the critical values of the form $\mathbf{x} =
\mathbf{x}_0 + \mathbf{u}$, where $\mathbf{x}=(x,y,z)$ and $\mathbf{u}
= (\delta x, \delta y, \delta z)$. The equations of
motion~(\ref{evolution eqs exp1}) can be written as $\mathbf{x}^\prime
= \mathbf{f}(\mathbf{x})$, which upon linearization reads
\begin{equation}
  \label{eq:linear}
  \mathbf{u}^\prime = \mathcal{M} \mathbf{u} \, , \quad \mathcal{M}_{ij}
  = \left. \frac{\partial f_i}{\partial x_j}
  \right|_{\mathbf{x}_\ast} \, , 
\end{equation}
where $\mathcal{M}$ is called the linearization matrix. The
eigenvalues $\omega$ of $\mathcal{M}$ determine the stability of the
critical points, whereas the eigenvectors $\mathbf{\eta}$ of
$\mathcal{M}$ determine the principal directions of the
perturbations. In general, if $\mathrm{Re}(\omega) < 0$
($\mathrm{Re}(\omega) > 0$) the critical point is called stable
(unstable).

In principle, we should study the perturbations of the three dynamical
variables $(x,y,z)$, but we should remember that they are not all
independent because they are bond together by the Friedmann
constraint~(\ref{Fconstriction}), and the same happens for their
perturbations.

The Friedmann constraint defines a two dimensional surface upon which
lie all physically relevant phase space trajectories, and then we will
be interested on perturbations lying also on the constraint surface. In
other words, perturbations which are perpendicular to the constraint
surface should be taken away from the analysis. 

We can identify the excluded perturbations by comparing their
associated eigenvectors with the gradient of the constraint surface at
each critical point,
\begin{eqnarray}
  \label{eq:gradient}
  \left. \nabla F \right|_{\mathbf{x}_\ast} &=& x_\ast \mathbf{i} +
    \left( y_\ast \pm \sqrt{(\lambda^2 -6)/3} z_\ast \right)
    \mathbf{j} \nonumber \\
    && + \left( z_\ast \pm \sqrt{(\lambda^2 -6)/3} y_\ast \right)
    \mathbf{k} \, .  
\end{eqnarray}
We will only take into account eigenvalues associated to eigenvectors
for which $\mathbf{\eta} \cdot \left. \nabla F
\right|_{\mathbf{x}_\ast} = 0$. The stability results are also
summarized in Table~\ref{tab:critical}.

Here we list the eigenvalues of the stability matrix $\mathcal{M}$ for
each of the critical points, only for the perturbations that are
compatible with the Friedmann constraint.
\begin{itemize}
\item Point A. It is an unstable point with eigenvalues
  \begin{equation}
    \label{eq:eigenwA}
    \omega_1 = 3 \, , \quad \omega_2 = 0 \, .
  \end{equation}
  The instability happens only along the eigenvector corresponding to
  $\omega_1$, which points in the positive $y$-direction. The second
  eigenvalue is null, and then the system is indifferent under
  perturbations along the unitary circumference $x^2+z^2=1$.

\item Point B. The stability eigenvalues for the cases $x_\ast = \mp
  1$ are
  \begin{equation}
    \label{eq:eigenwB}
    \omega_1 = \sqrt{\frac{3}{2}} (\lambda \pm \sqrt{6} ) \, , \quad
    \omega_2 = 0 \, .
  \end{equation}
  Similarly to the case of point A, only the eigenvector corresponding
  to $\omega_1$, which points in the positive $y$-direction, gives
  information about the stability of the critical points. We notice
  that the point at $x= -1$ is unstable, whereas that at $x=1$ can be
  unstable or saddle, depending upon the value of $\lambda$. Again,
  the second eigenvalue is null, and then the system is indifferent
  under perturbations along the unitary circumference $x^2+z^2=1$.

\item Points C. The eigenvalues corresponding to $x_\ast = \mp
  \sqrt{1-z^2_\ast}$ are
  \begin{equation}
    \label{eq:eigenwC}
    \omega_1 = 3 \pm \sqrt{\frac{3}{2}} \lambda \sqrt{1-z^2_\ast} \, ,
    \quad \omega_2 = 0 \, .
  \end{equation}
  These points have stability properties very similar to points A and B,
  but stability also depends upon their exact location on the unitary
  circumference $x^2+z^2 = 1$. 

  Notice that there is a special point,
  corresponding to $z_h = \sqrt{1-6/\lambda^2}$ and located at
  $(\sqrt{6}/\lambda,0,\sqrt{1-6/\lambda^2})$, which marks the
  instability-stability transition of the chain of points C. Because
  of this, there is a particular trajectory that departs from and also
  arrives to the point. This is called an homoclinic trajectory, and
  it is the largest loop that encloses point E, see
  Figs.~\ref{fig:fig1} and~\ref{fig:fig2}.

\item Point D. Their stability eigenvalues are
  \begin{equation}
    \label{eq:eigenwD}
    \omega_1 = -\frac{1}{2} ( 6 - \lambda^2 ) \, , \quad \omega_2 =
    - \frac{1}{2} ( 6 - \lambda^2 ) \, .
  \end{equation}
It is a stable point, whenever it exists, i.e. if the case $\lambda <
\sqrt{6}$ is allowed.

\item Point E. Its stability eigenvalues are
  \begin{equation}
    \label{eq:eigenwE}
    \omega_1 = i \sqrt{\frac{3}{2}} \frac{(\lambda^2 -
      6)^{3/2}}{\sqrt{9 - \lambda^2}} \, , \quad \omega_2 = -i
    \sqrt{\frac{3}{2}} \frac{(\lambda^2 -
      6)^{3/2}}{\sqrt{9 - \lambda^2}}\, .
  \end{equation}
The eigenvalues are purely imaginary for the range of existence of the
critical point, then it is a stable centre. This is confirmed by the
closed trajectories around the critical point in Fig.~\ref{fig:fig1}.
\end{itemize}

The overall conclusion is that only critical points A, B, C and E may
coexist together in the phase space, because the (inflationary) point
D is excluded by the (positivity) restriction $\lambda^2 > 6$. What we
observe in Figs.~\ref{fig:fig1} and~\ref{fig:fig2} is that the
ultimate fate for trajectories is to move around point E in closed
loops, for the case $6 < \lambda^2 < 9$, or to reach any of the stable
points C, for any $\lambda^2 > 6$. All possible solutions in the phase
space represent, in general, stiff matter solutions. 

\section{\label{sec:conclusions}Conclusions}
In this work we have considered a supersymmetric extension of the
action of general relativity for a scalar field interacting with the
scale factor of the Universe. For this purpose, we have introduced a
superfield formulation in which fermionic degrees of freedom are
associated to both the scale factor and to the scalar field.

By realizing the algebra of the fermionic variables and representing
them as matrices, we get four equations for four components of the wave
function. We focus our attention in two of them that are independent,
and apply the WKB method in order to get two \emph{classical}
SUSY-cosmological equations. The associated equations of motion for
the scalar field are obtained by means of Hamilton's equations.  

In these supersymmetric Eisntein-Klein-Gordon equations (SUSY-EKG),
new contributions arise that behave like stiff matter, and some others
in which the usual scalar field terms are modified by functions of the
scale factor. 

For simplicity, we focused our attention in a flat Universe and were
able to find exact solutions of the equations of motion. In one of
them the scalar field potential is a negative constant, and then the
radius of the Universe is a periodic function. A second exact solution
of interest corresponds to the case of an exponential
(super)potential. This is a scaling solution corresponding to stiff
fluid matter as revealed by the power law behaviour of the scale
factor.

We performed an analysis of the dynamical system structure of the
SUSY-EKG equations in order to find all relevant physical
solutions. Not surprisingly, we found, basically, the same solutions
that appear in the standard classical case. However, the stability and
existence properties of the solutions were strongly modified by the
supersymmetric corrections.

In general, we can say that all solutions show that the scale factor
and the scalar field degrees of freedom behave like in the case of
stiff matter domination, because the inflationary solution is
absent. This a consequence of the correction terms that appear in the
equations of motion, in which the scalar field functions are mediated
by stiff-matter terms of the scale factor.

All conclusions are consequences only of the supersymmetric nature of the
equations of motion, and the solutions are expected to be relevant in
the early stages of the Universe. We cannot foretell the consequences
that may arise in the case of a less exact supersymmetry, or even a
broken one. But these are possibilities that may modify again the
phase structure of the solutions and allow the existence of
inflationary solutions. This is research beyond the purposes of the
present work that we expect to report elsewhere.

\begin{acknowledgments}
CE-R acknowledges support from Fundaci\'on Pablo Garc\'ia, FUNDEC
M\'exico, and CONACyT M\'exico. This work was partially supported by
PROMEP, DAIP, and by CONACyT M\'exico under grants 56946, 51306, and
I0101/131/07 C-234/07 of the Instituto Avanzado de Cosmologia (IAC)
collaboration (http://www.iac.edu.mx/).
\end{acknowledgments}

\bibliography{susyreferences}

\end{document}